\documentclass[aip,apl,reprint,english,twocolumn,showpacs,preprintnumbers,amsmath,amssymb,floatfix]{revtex4-1}

\usepackage[T1]{fontenc}
\usepackage{graphicx}
\begin{document}

\title{Ultrathin Metallic Coatings Can Induce Quantum Levitation between Nanosurfaces}

\author{Mathias Bostr{\"o}m}
\email{mabos@ifm.liu.se}
\affiliation{Department of Energy and Process Engineering, Norwegian University of Science and Technology, N-7491 Trondheim, Norway}
\affiliation{Department of Applied Mathematics, Australian National University, Canberra, Australia}

\author{Barry W. Ninham}
\affiliation{Department of Applied Mathematics, Australian National University, Canberra, Australia}

\author{Iver Brevik}
\affiliation{Department of Energy and Process Engineering, Norwegian University of Science and Technology, N-7491 Trondheim, Norway}

\author{Clas Persson}
\affiliation{Dept of Materials Science and Engineering, Royal Institute of Technology, SE-100 44 Stockholm, Sweden}

\affiliation{Department of Physics, University of Oslo, P. Box 1048 Blindern,
NO-0316 Oslo, Norway}

\author{Drew F. Parsons}
\affiliation{Department of Applied Mathematics, Australian National University, Canberra, Australia}

\author{Bo E. Sernelius}
\email{bos@ifm.liu.se}
\affiliation{Division of Theory and Modeling, Department of Physics, 
Chemistry and Biology, Link\"{o}ping University, SE-581 83 Link\"{o}ping, Sweden}

\begin{abstract}
There is an attractive Casimir-Lifshitz force between two silica surfaces in a liquid (bromobenze or toluene). We demonstrate that adding an ultrathin (5-50\AA) metallic nanocoating to one of the surfaces results in repulsive Casimir-Lifshitz forces above a critical separation. The onset of such quantum levitation comes at decreasing separations as the film thickness decreases. Remarkably the effect of retardation can turn attraction into repulsion. From that we explain how an ultrathin metallic coating may prevent nanoelectromechanical systems from crashing together.
\end{abstract}


\maketitle

%\section{Introduction}

At close distances particles experience a Casimir-Lifshitz force (van der Waals force).\,\cite{ Lond, Casi,Dzya, Maha,Ser, Milt, Pars, Ninhb}
This takes a weaker (retarded) form with increasing separation.\,\cite{ Casi,Dzya}
We show how addition of ultrathin nanocoatings to interacting surfaces can change the forces from attractive to repulsive. This can be done by exploiting dielectric properties  to shift  the retarded regime down to a few nanometers. The addition of very thin coatings may also give repulsive van der Waals interactions (non-retarded Casimir-Lifshitz interactions) in asymmetric  situations. While these curious effects were in principle known 40 years ago\,\cite{Maha,Ninhb,Parsegian1971} they have not been explored in detail.  Nanotechnological advances now allow their exploitation.

\begin{figure}
\includegraphics[width=6cm]{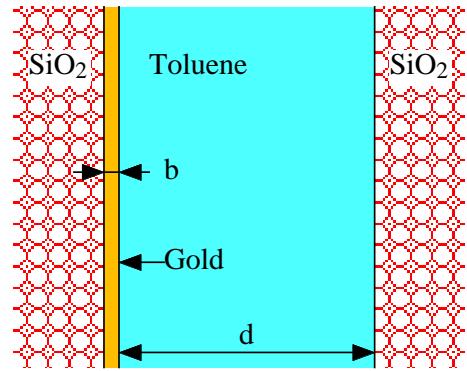}
\caption{ Model system where repulsive Casimir-Lifshitz interaction can be induced between silica surfaces in toluene when one surface has an ultrathin  gold nanocoating.}
\label{figu1}
\end{figure}

 In this letter the focus is on the interaction between gold coated silica and silica across toluene. For thick coating there is a retardation driven repulsion that sets in at {11 \AA} while for thin coating there can also be repulsion in the case when retardation is neglected.  There are in the case of one surface with thin gold coating and one bare surface not only a retardation driven repulsion\,\cite{bosserPRA2012} but also an effect due to the system being an asymmetric multilayer system.\,\cite{Maha} A schematic illustration of the system is shown in Fig. \ref{figu1}. 
Leaving one surface bare and treating one surface with an ultrathin metallic nanocoatings so provides  a way to induce what might be termed quantum levitation, i.e reduced friction  between equal surfaces in a liquid. The Casimir-Lifshitz energy depends in a very sensitive way on differences between dielectric functions.  When a {5-50 \AA} ultrathin gold  coating is added to one of the silica surfaces the force becomes repulsive.  It becomes repulsive for separations above a critical distance.\,\cite{bosserPRA2012}  This distance, which we will refer to as the levitation distance, decreases with decreasing film thickness. At large separation the interaction is repulsive; when separation decreases the repulsion increases, has a maximum, decreases and becomes zero at the levitation distance; for even smaller separation the interaction is attractive. The repulsion maximum increases with decreasing gold-film thickness.

\begin{figure}
\includegraphics[width=7.6cm]{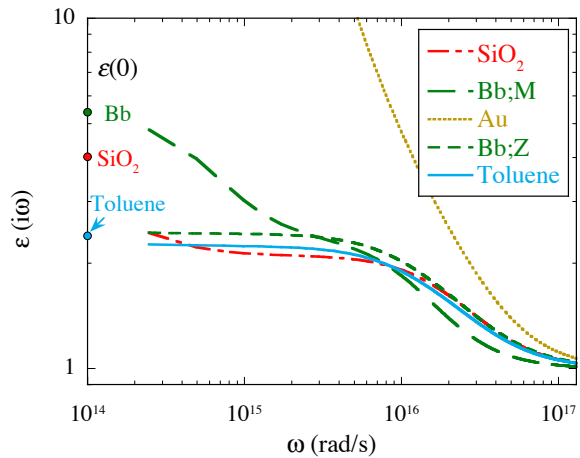}
\caption{ The dielectric function at imaginary frequencies for SiO$_2$ (silica)\,\protect \cite{Grab}, Bb (bromobenzene)\,\protect \cite{Mund,Zwol1}, Au (gold)\,\protect \cite{Ser}, and Toluene.\,\protect \cite{Zwol1} The static values have been displayed at the left vertical axis. }
\label{figu2}
\end{figure}

We study two silica  (SiO$_2$)\,\cite{Grab} surfaces in different liquids [bromobenzene (Bb) with data from Munday {\it et al.},\,\cite{Mund} bromobenzene with data from van Zwol and Palasantzas,\,\cite{Zwol1} and toluene\,\cite{Zwol1}]. 
To enable calculation of Casimir-Lifshitz energies a detailed knowledge of the dielectric functions is required.\,\cite{Dzya,Maha} Examples of such functions are shown in Fig.\,\ref{figu2}. As noted by Munday {\it et al.} \cite{Mund}  the fact that there is a crossing between the curves for SiO$_2$ and Bb opens up for the possibility of a transition of the Casimir-Lifshitz energy, from attraction to repulsion. A similar effect was seen earlier in another very subtle experiment performed by Hauxwell and Ottewill.\,\cite{Haux} They measured the thickness of oil films on water near the alkane saturated vapor pressure. For this system n-alkanes up to octane spread on water. Higher alkanes do not spread. It was an asymmetric system (oil-water-air) and the surfaces were molecularly smooth. The phenomenon depends on a balance of van der Waals forces against the vapor pressure.\,\cite{Rich, Haux, Ninh}

First we consider the model dielectric function for bromobenzene from Munday {\it et al.}\,\cite{Mund} also used in Ref.\,\onlinecite{bosserPRA2012}. This leads to the conclusion that retardation effects turn attraction into repulsion at the levitation distance. However, when we use the correct form for the dielectric function of bromobenzene from  van Zwol and Palasantzas\,\cite{Zwol1} the Casimir-Lifshitz force is repulsive also in the non-retarded limit! 
We will finally consider Casimir-Lifshitz interactions in toluene. This system provides us with an example where retardation turns attraction into repulsion for levitation distances of less than {11 \AA}. This is apparently counterintuitive as retardation effects are usually assumed to set in at much larger separations!  This is the case when the two interacting objects are immersed in vacuum. If they are immersed in a liquid or gas different frequency regions may give attractive and repulsive contributions. The net result depends on the competition between these attractive and repulsive contributions. It is obviously very important to have reliable data for the dielectric functions of the objects and ambient. 

 Quantum levitation from the Casimir effect modulated by thin conducting films may be a way to prevent surfaces used in quantum mechanical systems to come together by attractive van der Waals forces.  One important area for the application of van der Waals/Casimir theory is that of microelectromechanical systems (MEMS), as well as its further  extension NEMS (nanoelectromechanical systems). The demonstration of the first MEMS in the middle 1980's generated a large interest in the engineering community, but the practical usefulness of the technology has been much less than what was anticipated at its inception. One of the key barriers to commercial success has been the problem of stiction.\,\cite{serry1,serry2} Stiction is the tendency of small devices to stick together, and occurs when surface adhesion forces are stronger than the mechanical restoring force of the microstructure. The application of thin surface layers has turned out to be a  possibility to reduce or overcome the problem. In particular, as discussed in this letter, if the use of thin metallic layers creates an over-all repulsion between closely spaced surfaces in practical cases, this will be quite an attractive option.
Serry {\it et al.} have given  careful discussions of the relationship between the Casimir effect and stiction in connection with MEMS.\,\cite{serry1,serry2} The reader may  also consult the extensive and general review article  of Bushan.\,\cite{bhushan}

The field of measurements of quantum induced forces due to vacuum fluctuations was pioneered long ago by Deryaguin and Abrikossova.\,\cite{Der}  Lamoreaux\,\cite{Lamo} performed the first high accuracy measurement of Casimir forces between metal surfaces in vacuum that apparently confirmed predictions for both the Casimir asymptote and the classical asymptote.\,\cite{Sush,Milt2} The first measurements of Casimir-Lifshitz forces directly applied to MEMS were performed by Chan {\it et al.}\,\cite{ChanScience,ChanPRL} and somewhat later by Decca {\it et al.}.\,\cite{Decca} A key aspect of the Casimir-Lifshitz force is that according to theory it can  be either attractive or repulsive.\,\cite{Dzya,Rich1,Rich}  Casimir-Lifshitz repulsion was measured for films of liquid helium (10-200 \AA) on smooth surfaces.\,\cite{AndSab} The agreement found from theoretical analysis of these experiments meant a great triumph for the Lifshitz theory.\,\cite{Maha,Rich1,Haux,Ninh}  Munday, Capasso, and Parsegian\,\cite{Mund} carried out  direct force measurements showing that Casimir-Lifshitz forces could be repulsive. They found attractive Casimir-Lifshitz forces between gold surfaces in bromobenzene. When one surface was replaced with silica the force turned repulsive. It was recently shown that the repulsion may be a direct consequence of retardation.\,\cite{Phan11,bosserPRA2012} Only a few force measurements of repulsive Casimir-Lifshitz forces have been reported in the litterature.\,\cite{Mund,Milling,Lee,Feiler,Zwol1}

%\section{Theory and Numerical Results}

Now to the actual calculations and numerical results. One way to find retarded van der Waals or Casimir-Lifshitz interactions is in terms of the electromagnetic normal modes\cite{Ser} of the system. For planar structures the interaction energy per unit area can be written as
%1
\begin{equation}
E = \hbar \int {\frac{{{d^2}k}}{{{{\left( {2\pi } \right)}^2}}}} \int\limits_0^\infty  {\frac{{d\omega }}{{2\pi }}} \ln \left[ {{f_k}\left( {i\omega } \right)} \right],
\label{equ1}
\end{equation}
where ${f_k}\left( {{\omega _k}} \right) = 0$ is the condition for electromagnetic normal modes. Eq.\,(\ref{equ1}) is valid for zero temperature and the interaction energy is the internal energy. At finite temperature the interaction energy is Helmholtz' free energy and can be written as
%2
\begin{equation}
\begin{array}{l}
E = \frac{1}{\beta }\int {\frac{{{d^2}k}}{{{{\left( {2\pi } \right)}^2}}}} \sum\limits_{n = 0}^\infty {'} {\ln \left[ {{f_k}\left( {i{\omega _n}} \right)} \right];} \\
{\omega _n} = \frac{{2\pi n}}{{\hbar \beta }};\;n = 0,\,1,\,2,\, \ldots ,
\end{array}
\label{equ2}
\end{equation}
where $\beta  = 1/{k_B}T$. The integral over frequency has been replaced by a summation over discrete Matsubara frequencies. The prime on the summation sign indicates that the $n = 0$ term should be divided by two. 
For planar structures the quantum number that characterizes the normal modes is {\bf k}, the two-dimensional (2D) wave vector in the plane of the interfaces and there are two mode types, transverse magnetic (TM) and transverse electric (TE).

The general expression for the mode condition function for two coated planar objects in a medium, i.e., for the geometry 1|2|3|4|5 is
\begin{equation}
{f_k} = 1 - {e^{ - 2{\gamma _3}k{d_3}}}{r_{321}}{r_{345}},
\label{equ3}
\end{equation}
where\,\cite{Maha, Ser} 
\begin{equation}
{r_{ijk}} = \frac{{{r_{ij}} + {e^{ - 2{\gamma _j}k{d_j}}}{r_{jk}}}}{{1 + {e^{ - 2{\gamma _j}k{d_j}}}{r_{ij}}{r_{jk}}}},
\label{equ4}
\end{equation}
and
\begin{equation}
{\gamma _i} = \sqrt {1 - {\varepsilon _i}\left( \omega  \right){{\left( {\omega /ck} \right)}^2}}.
\label{equ5}
\end{equation}
The function ${{\varepsilon _i}\left( \omega  \right)}$ is the dielectric function of medium $i$. The amplitude reflection coefficients for a wave impinging on an interface between medium $i$ and $j$ from the $i$-side are
\begin{equation}
r_{ij}^{TM} = \frac{{{\varepsilon _j}{\gamma _i} - {\varepsilon _i}{\gamma _j}}}{{{\varepsilon _j}{\gamma _i} + {\varepsilon _i}{\gamma _j}}},
\label{equ6}
\end{equation}
and
\begin{equation}
r_{ij}^{TE} = \frac{{\left( {{\gamma _i} - {\gamma _j}} \right)}}{{\left( {{\gamma _i} + {\gamma _j}} \right)}},
\label{equ7}
\end{equation}
for TM and TE modes, respectively.

In the present work we  calculate the Casimir-Lifshitz energy between a gold coated silica surface and a silica surface across a liquid, i.e. we study the geometry 1|2|3|1, where medium  3 is the liquid. The mode condition function for this geometry is ${f_k} = 1 - {e^{ - 2{\gamma _3}k{d_3}}}{r_{321}}{r_{31}}$.

\begin{figure}
\includegraphics[width=7.6cm]{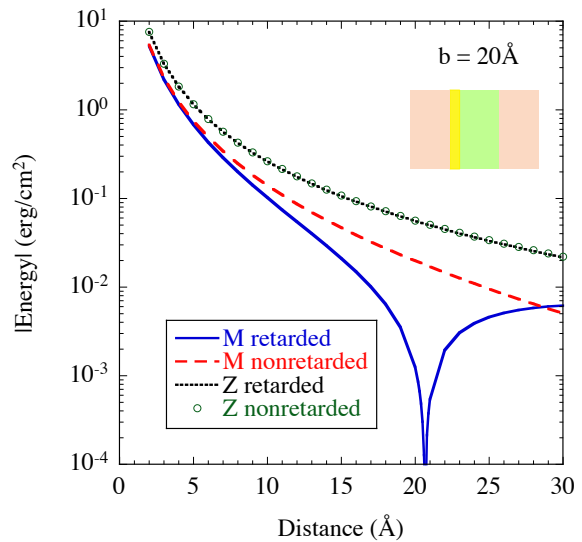}
\caption{ The retarded and non-retarded Casimir-Lifshitz interaction free energy between a silica surface and a gold coated (b = 20 \AA) silica surface in bromobenzene using different dielectric functions for bromobenzene from Munday {\it et al.} (M)\,\protect \cite{Mund} and from van Zwol {\it et al.} (Z).\,\protect \cite{Zwol1} The result of the Munday model gives repulsion only in the retarded treatment. In contrast the van Zwol model gives repulsion in both the retarded and non-retarded limits.}
\label{figu3}
\end{figure}

We now demonstrate in Fig.\,\ref{figu3} how different models for the dielectric function of bromobenzen (given by Munday {\it et al.}\,\cite{Mund} and by van Zwol {\it et al.}\,\cite{Zwol1}) produce fundamentally different results for the role played by retardation in the repulsive Casimir-Lifshitz force. The difference between the two models is that  van Zwol and Palasantzas\,\cite{Zwol1} treated the contributions from lower frequency ranges in a more accurate way.  The prediction using the model from van Zwol {\it et al.}\,\cite{Zwol1} is that the interaction is repulsive also when retardation is not accounted for.  This is in contrast to a retardation driven repulsion found with the data given by the model from Munday {\it et al.}\,\cite{Mund}

\begin{figure}
\includegraphics[width=7.6cm]{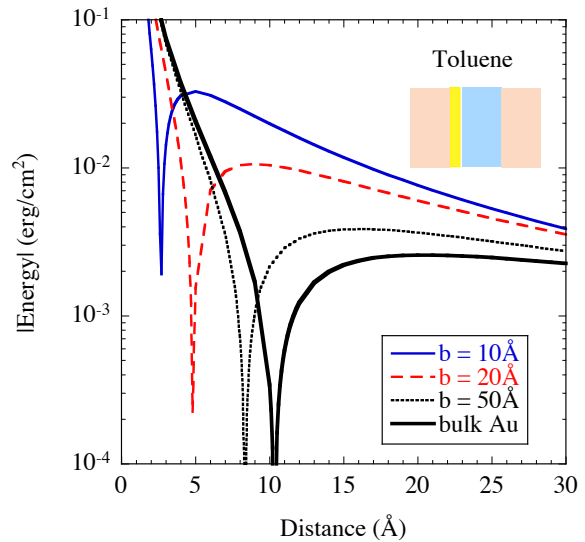}
\caption{ The retarded Casimir-Lifshitz interaction free energy between a silica surface and a gold coated silica surface in toluene using dielectric function for  toluene from van Zwol {\it et al.}\,\protect \cite{Zwol1} The interaction is attractive at short distances and repulsive above a critical levitation distance.}
\label{figu4}
\end{figure}
\begin{figure}
\includegraphics[width=7.6cm]{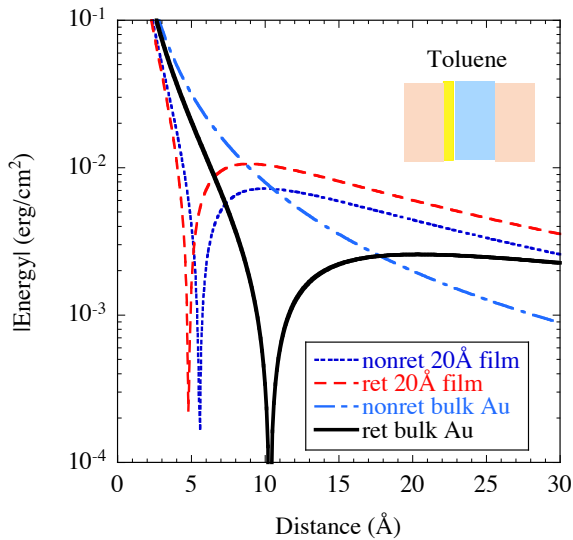}
\caption{ The retarded and nonretarded Casimir-Lifshitz interaction free energy between a silica surface and a gold coated silica surface in toluene using dielectric function for  toluene from van Zwol {\it et al.}\,\protect \cite{Zwol1} The nonretarded interaction between thick gold films and silica across toluene is attractive for all distances. The other examples considered (nonretarded and retarded for the case of {20 \AA} gold film and retarded with thick gold films) all cross over to repulsion above a critical distance.}
\label{figu5}
\end{figure}

To finish up we present in Fig.\,\ref{figu4}  what appears to be a very promising system for studying retardation effect for very small separations: gold coated silica interacting with silica in toluene. Here the levitation distance comes in the range from a few {\AA ngstr{\"o}ms} up to {11 \AA} for thick gold films.  For a gold surface interacting with silica across toluene the non-retarded Casimir-Lifshitz force is attractive for all separations. This suggests that it  is possible to  have repulsion in the nanometer range induced by metal coatings and retardation. We show in Fig.\,\ref{figu5} the nonretarded and retarded Casimir-Lifshitz interaction energies between two silica surfaces in toluene when one of the surfaces has a {20 \AA} gold nanocoating  or a very thick gold coating. Here it is more evident that there are two effects that combine to give repulsion at very small distances: the finite thickness of the film  (which by itself leads to repulsive van der Waals interaction energies) and retardation. The enhancement of the repulsive Casimir-Lifshitz energy for thin films as compared to thick films is then seen to be mainly related to the finite film thickness and to a lesser degree to retardation.

%\section{Conclusions}

To conclude, we have seen that the effects of retardation turn up already at distances of the order of a few nm or less.  Remarkably the effect of retardation can be to turn attraction into repulsion in a way that depends strongly on the optical properties of the interacting surfaces. Addition of ultrathin metallic coatings may prevent nanoelectromechanical systems from crashing together. Quantum levitation from addition of ultrathin conducting coatings may provide a well needed revitalization of the field of MEMS and NEMS.   As pointed out by Palasantzas and co-workers\,\cite{Zwol1} it is crucial to obtain accurate dielectric functions from optical data or from calculations. The exact levitation distances vary with choice of dielectric functions. It is important to use accurate optical data to be able to correctly predict levitation distances for specific combinations of materials.

%\begin{acknowledgments}
 MB acknowledge support from an European Science Foundation exchange grant within the activity "New Trends and Applications of the Casimir Effect" through the network CASIMIR. B.E.S. acknowledge financial support from VR (Contract No. 70529001). 
%\end{acknowledgments}

\end{document}